\documentclass{elsart}
\usepackage{epsfig,amssymb}

\begin{document}

\begin{frontmatter}
\title{\boldmath Spin dynamics in a structurally ordered non-Fermi liquid compound: YbRh$_2$Si$_2$}
\author[UCR,OU]{K. Ishida,\thanksref{KU}}
\author[UCR]{\underline{D. E. MacLaughlin},}
\author[CSU]{O. O. Bernal,}
\author[LANL]{R. H. Heffner,}
\author[KOL]{G. J. Nieuwenhuys,}
\author[MPI]{O. Trovarelli,}
\author[MPI]{C. Geibel} and
\author[MPI]{F. Steglich}
\thanks[KU]{Present address: Department of Physics, Kyoto University, Kyoto 606-8502, Japan.}

\address[UCR]{Department of Physics, University of California, \\Riverside, California 92521, USA}
\address[OU]{Department of Physical Science, Graduate School of Engineering Science, \\Osaka University, Toyonaka, Osaka 560-8531, Japan}
\address[CSU]{Department of Physics and Astronomy, California State University,\\ Los Angeles, California 90032, USA}
\address[LANL]{MS K764, Los Alamos National Laboratory, \\Los Alamos, New Mexico 87545, USA}
\address[KOL]{Kamerlingh Onnes Laboratorium, Leiden University, \\2300 RA Leiden, The Netherlands}
\address[MPI]{Max-Planck Institute for Chemical Physics of Solids, D-01187 Dresden, Germany}

\begin{abstract}
Muon spin relaxation ($\mu$SR) experiments have been carried out at low temperatures in the non-Fermi-liquid heavy-fermion compound YbRh$_2$Si$_2$. The longitudinal-field $\mu$SR relaxation function is exponential, indicative that the dynamic spin fluctuations are homogeneous. The relaxation rate $1/T_1$ varies with applied field as $H^{-y}$, $y = 1.0 \pm 0.1$, which implies a scaling law of the form~$\chi''(\omega) \propto \omega^{-y} f(\omega/T)$, $\lim_{x\to0} f(x) = x$ for the dynamic spin susceptibility.

\end{abstract}
\begin{keyword}
Non-Fermi liquids, quantum criticality, heavy-fermion compounds, YbRh$_2$Si$_2$.
\end{keyword}
\end{frontmatter}

Non-Fermi liquid (NFL) behavior and quantum critical (QC) phenomena in strongly correlated electron systems are among the most intensively studied subjects in condensed matter physics~\cite{ITP96els}. NFL behavior is observed in a number of f-electron systems as pronounced deviations from conventional Landau Fermi-liquid properties. These include anomalous temperature dependences of the electrical resistivity [$\Delta \rho = \rho(T) - \rho(T{=}0) \propto T^{\alpha}$, $1 \lesssim \alpha < 2$] and the f-electron specific heat coefficient ($C_{\rm el} / T \propto -\ln T$)~\cite{Stew01,Steg01}. This contrasts with the Fermi-liquid behavior seen in ordinary metals ($\Delta \rho \propto T^2$, $C_{\rm el}/T = {\rm const.}$). One mechanism for NFL behavior invokes a quantum critical point (QCP), which is attained by varying a control parameter such as doping, pressure, or magnetic field~\cite{Stew01}.

Microscopic techniques such as $\mu$SR are powerful tools with which to investigate QC magnetic fluctuations. NMR and $\mu$SR experiments on the NFL alloys~UCu$_{5-x}$Pd$_x$, $x = 1.0$ and 1.5, have revealed strongly inhomogeneous spin fluctuations and glassy dynamics; the $\mu$SR results suggest a quantum spin-glass state near the QCP~\cite{MBHN01} with extremely slow fluctuations, i.e., thermally-excited fluctuations with strong spectral weight at very low frequencies. Thus disorder may strongly influence the spin fluctuation behavior, so that it is important to determine whether or not low-frequency fluctuations associated with a QCP are present in ordered stoichiometric compounds.

The NFL compound~YbRh$_2$Si$_2$ appears to be a suitable system for the study of such ``clean'' NFL physics. YbRh$_2$Si$_2$ appears to be located in the vicinity of a QCP from a number of bulk measurements~\cite{TGML00}; resistivity and specific heat measurements at low temperatures show $\Delta \rho \propto T$ and $C_{\rm el} / T \propto -\ln T$ over a temperature range of more than a decade. The NFL behavior is suppressed by the application of an external magnetic field, and Fermi-liquid behavior is recovered. Low-field ac susceptibility measurements show an anomaly suggestive of a magnetic phase transition around 70 mK that is suppressed by small magnetic field of $\sim$500 Oe. These results suggest that YbRh$_2$Si$_2$ is quite close to a QCP, but undergoes a phase transition at very low temperatures and magnetic fields.

In order to investigate the 70-mK anomaly and the character of spin fluctuations at zero and weak magnetic fields, we have carried out muon spin relaxation experiments on YbRh$_2$Si$_2$ at the LTF facility of the Paul Scherrer Institute, Switzerland. A full report of this work, which will be published elsewhere~\cite{IMOK02b}, discusses in more detail questions of static magnetism in the low-temperature state in YbRh$_2$Si$_2$, the muon stopping site, and static susceptibility inhomogeneity.

\begin{figure}
\begin{center}
\epsfig{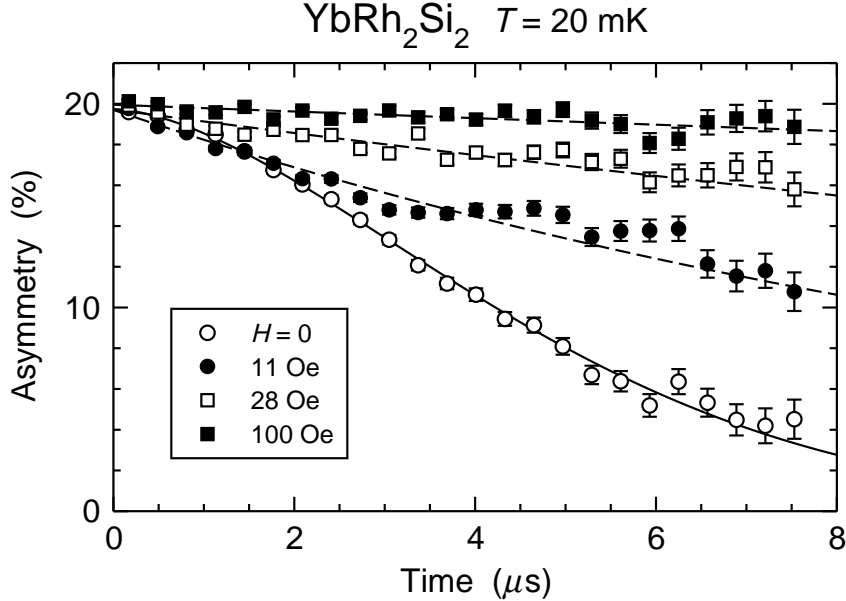}
\end{center}
\caption{LF-$\mu$SR relaxation in YbRh$_2$Si$_2$, $T = 20$ mK, longitudinal field~$H = 0$, 11, 28, and 100 Oe. Solid curve: fit to static Kubo-Toyabe relaxation function for $H = 0$. Dashed curves: fits to exponential relaxation function for $H > 0$.}
\label{fig:YbRhSifig1}
\end{figure}

Figure~\ref{fig:YbRhSifig1} shows the relaxation function~$G(t)$ for the muon decay asymmetry in YbRh$_2$Si$_2$ at $T = 20$ mK for longitudinal fields~$H = 0$, 11, 28, and 100 Oe. 
At zero field the data can be fit to a static Kubo-Toyabe (K-T) relaxation function \cite{HUIN79} corresponding to a random distribution of static $\mu^+$ local fields with rms $\mathrm{width} \approx 2$ Oe. As discussed elsewhere \cite{IMOK02b,IMOK02a}, we attribute these static fields to weak ordered Yb moments ($10^{-3}$--$10^{-2}\ \mu_B$) in the antiferromagnetic state (N\'eel temperature~$T_N = 70$ mK), since the K-T rate is much larger than expected from nuclear dipolar fields and sets in only below $T_N$. The relaxation data for nonzero applied field can be fit to an exponential relaxation function~$G(t) = A\exp(-t/T_1)$. An applied field of 11 Oe is more than five times larger than the estimated field at the muon site due to static Yb-moment magnetism~\cite{IMOK02b}, and therefore is large enough to decouple this static field. Thus the relaxation observed for $H \gtrsim 10$ Oe is dynamic. Therefore the field dependence of $1/T_1$ is not due to decoupling, but instead suggests a significant frequency dependence to the local-field fluctuation spectrum at the low muon frequencies. 

It is important to note that the observed exponential form is evidence that the relaxation rate~$1/T_1$ is substantially uniform throughout the sample, since the signature of inhomogeneity in $1/T_1$ is a sub-exponential or ``stretched'' exponential relaxation function~\cite{MBHN01}. The muon relaxation is therefore probing spin fluctuations in a structurally ordered non-Fermi liquid. 

\begin{figure}
\begin{center}
\epsfig{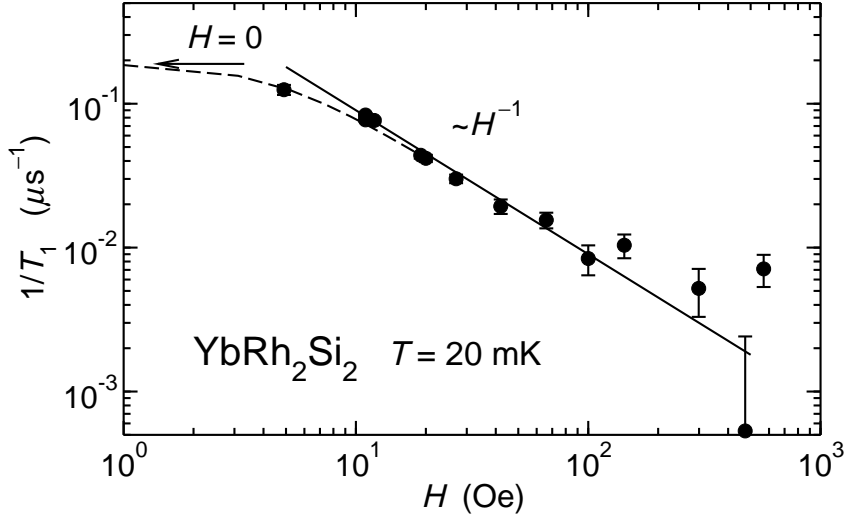}
\end{center}
\caption{Longitudinal field dependence of LF-$\mu$SR relaxation rate~$1/T_1$ in YbRh$_2$Si$_2$, $T = 20$ mK. Arrow: zero-field static K-T rate from fit of Fig.~\protect\ref{fig:YbRhSifig1}. The muon decay lifetime limits accuracy for $1/T_1 \lesssim 0.01\ \mu{\rm s}^{-1}$.}
\label{fig:YbRhSifig2}
\end{figure}

Figure~\ref{fig:YbRhSifig2} shows the dependence of $1/T_1$ on longitudinal field at $T = 20$ mK.
\suppressfloats 
The zero-field rate given by the arrow is the static K-T value obtained from the fit shown in Fig.~\ref{fig:YbRhSifig1}. The relaxation rate shows a weak field dependence for magnetic fields less than 10 Oe but varies more strongly, as $H^{-y}$, $y = 1.0 \pm 0.1$, for higher fields. 

In the motional narrowing limit $1/T_1$ is related to the spin autocorrelation function~$q(t) = \langle S_i(t)S_i(0)\rangle$ by~\cite{MBHN01,KMCL96} 
\begin{equation}
1/T_1(\omega,T) = \gamma_\mu^2\, \overline{\delta B^2}\int_0^\infty dt\,q(t)\cos\omega t \equiv \gamma_\mu^2\, \overline{\delta B^2}\, \tau_c(\omega,T) \,,
\end{equation}
where $\gamma_\mu$ is the muon gyromagnetic ratio, $\overline{\delta B^2}$ is the mean-square fluctuating field at the muon site, and $\tau_c(\omega,T)$ is the effective correlation time measured at the Zeeman frequency~$\omega = \gamma_\mu H$. Furthermore $\tau_c$ is related to the imaginary part of the complex susceptibility~$\chi''(\omega,T)$ by the fluctuation-dissipation theorem, which for $\hbar\omega \ll k_B T$ becomes
\begin{equation}
\tau_c(\omega,T) \propto (T/\omega) \chi''(\omega,T) \,.
\end{equation}
The observed frequency dependence of $1/T_1$ is therefore consistent with a scaling law of the form
\begin{equation}
\chi''(\omega,T) \propto \omega^{-y} f(\omega/T) \,,
\label{eq:chi}
\end{equation}
provided that $f(x) \to x$ as $x \to 0$. This scaling would then leave $1/T_1(\omega,T)$ independent of temperature, in rough agreement with the observed weak ($-\ln T$) dependence at $H = 19$ Oe (not shown)~\cite{IMOK02b}. For $\chi''(\omega)$ of the form of Eq.~(\ref{eq:chi}) this gives $q(t) \propto t^{-1+y}$ at long times \cite{KMCL96}. Thus $q(t)$ varies slowly with $t$, i.e., the low-temperature spin fluctuations are very long lived.

The exponential behavior of the relaxation function in the present $\mu$SR experiments indicates that YbRh$_2$Si$_2$ is an ordered stoichiometric compound. Thus disorder-driven theories~\cite{MDK96,CNJ00} seem to be ruled out, although these have been considered as promising scenarios for NFL behavior in disordered heavy-fermion materials. Our results strongly suggest that YbRh$_2$Si$_2$ is a compound in which NFL behavior is induced by homogeneous critical spin fluctuations.

We are grateful to A. Amato, C. Baines, and D. Herlach for help with the experiments, and to Q. Si for useful discussions. This work was supported in part by the U.S. NSF, Grant nos.~DMR-0102293 (UC Riverside) and DMR-9820631 (CSU Los Angeles), by the COE Research Grant-in-Aid (10CE2004) for Scientific Research from the Ministry of Education, Sport, Science, and Technology of Japan, and by the Netherlands NWO and FOM, and was performed in part under the auspices of the U.S. DOE.

\end{document}